\documentclass[a4paper,12pt, epsfig]{article}
\usepackage{epsfig}
\usepackage{amssymb}
\usepackage{amsfonts}
\usepackage{amsmath}
\usepackage{color}

\newskip\humongous \humongous=0pt plus 1000pt minus 1000pt

\newif\ifdtup

\catcode`\@=11

\@addtoreset{equation}{section}
\def\theequation{\thesection.\arabic{equation}}

\def\@normalsize{\@setsize\normalsize{15pt}\xiipt\@xiipt
\abovedisplayskip 14pt plus3pt minus3pt%
\belowdisplayskip \abovedisplayskip
\abovedisplayshortskip \z@ plus3pt%
\belowdisplayshortskip 7pt plus3.5pt minus0pt}

\def\small{\@setsize\small{13.6pt}\xipt\@xipt
\abovedisplayskip 13pt plus3pt minus3pt%
\belowdisplayskip \abovedisplayskip
\abovedisplayshortskip \z@ plus3pt%
\belowdisplayshortskip 7pt plus3.5pt minus0pt
\def\@listi{\parsep 4.5pt plus 2pt minus 1pt
     \itemsep \parsep
     \topsep 9pt plus 3pt minus 3pt}}

\relax

\catcode`@=12

\catcode`\@=11

\def\section{\@startsection{section}{1}{\z@}{3.5ex plus 1ex minus
   .2ex}{2.3ex plus .2ex}{\large\bf}}

\def\thesection{\arabic{section}}
\def\thesubsection{\arabic{section}.\arabic{subsection}}

\def\appendix{\setcounter{section}{0}
 \def\thesection{Appendix \Alph{section}}
 \def\thesubsection{\Alph{section}.\arabic{subsection}}
 \def\theequation{\Alph{section}.\arabic{equation}}}
\def\SymBoxes#1#2#3#4{\newdimen\un@t \un@t#3%
\raisebox{#1}{\rule{#2\un@t}{#4}\hskip-#2\un@t% lower horizontal
\@tempdimb\un@t \advance\@tempdimb by-#4\@tempcntb#2\relax%
\@whilenum{\@tempcntb>0}\do{%                         % #2 vertical lines
\rule{#4}{\un@t}\hskip\@tempdimb \advance\@tempcntb by\m@ne}%
\hskip-#2\un@t \rule[\un@t]{#2\un@t}{#4}%
\rule[\un@t]{#4}{#4}\hskip-#4%             % upper horizontal line
\rule{#4}{\un@t}}\hskip-#4}                % rightest vertical line
\begin{document}
%\begin{letter}{~}

\newcommand{\pa}{\partial}

%%%%%%Define some new commands and  macros
\newcommand{\beq}{\begin{equation}}
\newcommand{\eeq}{\end{equation}}
\newcommand{\bea}{\begin{eqnarray}}
\newcommand{\eea}{\end{eqnarray}}
\newcommand{\beas}{\begin{eqnarray*}}
\newcommand{\eeas}{\end{eqnarray*}}
\newcommand{\defi}{\stackrel{\rm def}{=}}
\newcommand{\non}{\nonumber}
\newcommand{\bquo}{\begin{quote}}
\newcommand{\enqu}{\end{quote}}
%%%%%%%%%%%%%%%%
\renewcommand{\(}{\begin{equation}}
\renewcommand{\)}{\end{equation}}
%%%%%%%%%%%%%%%%%%%%%%%%%%%%%%%%%% definitions
\def\IZ{{\mathbb Z}}
\def\IR{{\mathbb R}}
\def\IC{{\mathbb C}}
\def\IQ{{\mathbb Q}}

\def\g{\gamma}
\def\m{\mu}
\def\n{\nu}
\def\a{\alpha}
\def\b{\beta}

\def\CM{{\mathcal{M}}}
\def\dCM{{\left \vert\mathcal{M}\right\vert}}

\def \d{\textrm{d}}
\def \p{\partial}

\def \Pf{\rm Pf\ }

\def \pr{\prime}

\def\Tr{ \hbox{\rm Tr}}
\def\half{\frac{1}{2}}

\def \eqn#1#2{\begin{equation}#2\label{#1}\end{equation}}
\def\de{\partial}
\def\Tr{ \hbox{\rm Tr}}
\def\H{ \hbox{\rm H}}
\def\HE{ \hbox{$\rm H^{even}$}}
\def\HO{ \hbox{$\rm H^{odd}$}}
\def\K{ \hbox{\rm K}}
\def\Im{ \hbox{\rm Im}}
\def\Ker{ \hbox{\rm Ker}}
\def\const{\hbox {\rm const.}}
\def\o{\over}
\def\im{\hbox{\rm Im}}
\def\re{\hbox{\rm Re}}
\def\bra{\langle}\def\ket{\rangle}
\def\Arg{\hbox {\rm Arg}}
\def\Re{\hbox {\rm Re}}
\def\Im{\hbox {\rm Im}}
\def\exo{\hbox {\rm exp}}
\def\diag{\hbox{\rm diag}}
\def\longvert{{\rule[-2mm]{0.1mm}{7mm}}\,}
\def\a{{\textsl a}}
\def\dag{{}^{\dagger}}
\def\tq{{\widetilde q}}
\def\p{{}^{\prime}}
\def\W{W}
\def\N{{\cal N}}
\def\hsp{,\hspace{.7cm}}
\newcommand{\C}{\ensuremath{\mathbb C}}
\newcommand{\Z}{\ensuremath{\mathbb Z}}
\newcommand{\R}{\ensuremath{\mathbb R}}
\newcommand{\rp}{\ensuremath{\mathbb {RP}}}
\newcommand{\cp}{\ensuremath{\mathbb {CP}}}
\newcommand{\vac}{\ensuremath{|0\rangle}}
\newcommand{\vact}{\ensuremath{|00\rangle}}
\newcommand{\oc}{\ensuremath{\overline{c}}}

\def\M{\mathcal{M}}
\def\F{\mathcal{F}}
\def\d{\textrm{d}}

\def\eps{\epsilon}

\begin{flushright}
%SITP-09/05,
%ITEP-TH-06/09
\end{flushright}

%\hfill
%\vspace{18pt}
\begin{center}
{\Large \textbf{Scale-invariant breaking of conformal symmetry}}
\end{center}

%\vspace{6pt}
\begin{center}
{\large\textsl{Anatoly Dymarsky $^{a_{\, ,}}$\footnote{Permanent address.} and Alexander Zhiboedov $^b$ }\\}
\vspace{15pt}
\textit{\small $^a$ Center for Theoretical Physics,
Massachusetts Institute of Technology, \\ Cambridge, MA 02139 }\\[5pt]
\textit{\small $^1$ Skolkovo Institute of Science and Technology, \\
Novaya St. 100, Skolkovo, Moscow Region, Russia, 143025}\\[5pt]
\textit{\small $^b$ Center for the Fundamental Laws of Nature,
Harvard University, \\ Cambridge, MA 02138 USA}
\end{center}

\vspace{12pt}

%\begin{center}
\noindent
\textbf{Abstract}\ \ 

Known examples of unitary relativistic scale but not conformal-invariant field theories (SFTs) can be embedded into conventional conformal field theories (CFTs). We show that any SFT which is a subsector of a unitary CFT is a free theory. Our discussion applies to an arbitrary number of spacetime dimensions and explains triviality of known SFTs in four spacetime dimensions. We comment on examples of unitary SFTs which are not captured by our construction.

%\vspace{4pt} 
%{\small

%\noindent }

%\vspace{1cm}

\renewcommand{\thefootnote}{\arabic{footnote}}
\section{Introduction}
One of the fundamental questions of quantum field theory is to understand the structure of fixed points of the renormalization group (RG) flow. It is a central question for the program of describing the landscape of quantum field theories. Further importance of this question stems from the relation between  fixed points of the RG flow and critical phenomena, and through AdS/CFT correspondence to quantum gravity. 

By definition scale invariance is the only characteristic feature of a fixed point. Remarkably, with a few known exceptions discussed below, unitary scale-invariant relativistic field theories always exhibit full conformal symmetry.  The mechanism behind symmetry enhancement remains poorly understood. Our current level of understanding links emergence of conformal symmetry  to irreversibility of the RG flow. This is certainly the case in two dimensions, where Zamolodchikov's $c$-function \cite{Zamolodchikov:1986gt} can be used to establish conformality of any unitary scale-invariant field theory with a well-defined stress-energy tensor \cite{Polchinski:1987dy}. Similarly, in four dimensions irreversibility of the RG flow, reflected by $a$-theorem \cite{Cardy:1988cwa, Komargodski:2011vj},  could be used to establish that the stress-energy tensor of a unitary scale-invariant field theory admits an improvement to a traceless form \cite{Dymarsky:2013pqa, Dymarsky:2014zja, Yonekura:2014tha}.  This means the SFT in question is either conformal or can be embedded into conformal field theory as a sub-sector.
Far less is known about scale-invariant theories in other dimensions. We would like to emphasize that there are known examples of scale-invariant theories in $d\neq 2, 4$ which can not be embedded into unitary CFTs, indicating situation   in other dimensions could be rather different. 

An intriguing feature of known scale-invariant but not conformal field theories is that these are theories of free fields. Sometimes this leads to an erroneous notion that these theories are not physical and somehow should be discarded, giving rise to an overly simplified picture that scale invariance plus unitarity always implies conformaltiy. Certainly this is not true as the free field theories provide a well-defined example of  non-conformal unitary SFTs. At the same time the reason why all known examples of non-conformal SFTs are free {\bf is } not well-understood and merits separate consideration.  This paper makes a first step in this direction
proving that any scale-invariant but not conformally-invariant sub-sector of a unitary conformal field theory in any number of dimensions is a free theory. Paraphrasing our result we show that it is pointless to look for interacting non-trivial SFTs within some intricate or complicated unitary CFT.  It does not matter which complicated or exotic  CFT we consider, there is no SFTs under the CFT lamppost.

\section{Scale invariance within conformality}\label{CFT}
Our starting point is a unitary conformal field theory in $d\ge 3$ dimensions. We assume the theory has a unique conserved traceless spin-two current, the stress-energy tensor,
\bea
\partial_\mu T_{\mu\nu}^{\rm CFT}=0\ ,\quad 
T^{\rm CFT}_{\mu\mu}=0\ .
\eea 
The generators of global conformal symmetry $P_\mu, D, K_\mu$ can be expressed as integrals of $T_{\mu\nu}^{CFT}$. 

We are interested in CFTs which include SFT as a sub-sector. For the sub-sector to be a well-defined  theory it must be closed under OPE algebra. This would be the case for the sub-sector invariant under a symmetry $S$.\footnote{In the following we assume $S$ is a generator of a continuous symmetry. Generalization to the discreet case is straightforward.} In other words we define SFT through the set of correlation functions of all operators $\mathcal O$ invariant under $S$, $[S,{\mathcal O}]=0$. 
Symmetry $S$ commutes with the Hamiltonian and by Lorentz invariance with all momenta $ [S,P_\mu]=0$. Scale-invariance of would-be SFT requires $S$ to be an eigenvector of dilatation
\bea
\label{SD}
[S,D]=\kappa S
\eea
for some fixed $\kappa$. In case there is more than one symmetry generator $S_I$ and SFT is a sector invariant under all symmetries, one can always find a combination of $S_I$ satisfying \eqref{SD}. Finally, the commutator $[S,K_\mu]$ can not be zero or proportional to $S$. Otherwise the scale-invariant sub-sector  would exhibit full conformal symmetry. 

Our definition of a well-defined scale-invariant theory assumes the theory includes a local conserved stress-energy tensor, 
\bea
\label{annihilation}
[S,T^{\rm SFT}_{\mu\nu}]=0\ .
\eea
Clearly $T^{SFT}_{\mu\nu}$ must differ from $T_{\mu\nu}^{CFT}$ by an improvement term \cite{Polchinski:1987dy},
\bea
\label{improv}
T_{\mu\nu}^{\rm CFT}=T_{\mu\nu}^{\rm SFT}+{1\over 2(d-1)}\left(\partial_\mu \partial_\alpha L_{\nu \alpha}+\partial_\nu \partial_\alpha L_{\mu \alpha} - \partial^2 L_{\mu\nu}-g_{\mu\nu} \partial_\alpha \partial_\beta L_{\alpha\beta}\right)\ .
\eea
It is convenient to decompose $L_{\mu\nu}$ into a scalar and traceless part 
\bea
L_{\mu\nu}=g_{\mu\nu}L+L_{\mu\nu}^{\perp}\ ,\quad L_{\mu\mu}^{\perp}=0\ .
\eea
Scale invariance dictates that $L^{\perp}_{\mu\nu}$ has dimension $d-2$.
In unitary CFTs the dimension of a traceless symmetric conformal primary of spin $\ell$ is bound 
by 
\bea
\Delta\ge d-2+\ell\ .
\eea Hence $L^{\perp}_{\mu\nu}$ can not be a primary or a descendant of an operator with a non-zero spin. The only possibility if $L^{\perp}_{\mu\nu}$ is a descendant of a primary scalar of dimension $\Delta_Y=d-4$, 
\bea
L^{\perp}_{\mu\nu}=\left(\partial_\mu \partial_\nu -{g_{\mu\nu}\over d} \partial^2  \right)Y\ ,
\eea
which requires $d\ge 6$. 

Since $T^{CFT}_{\mu\nu}$ is traceless, \eqref{improv} equates the trace of $T^{SFT}_{\mu\nu}$ with the Laplacian of some scalar operator, 
\bea
\label{trace}
T_{\mu\mu}^{\rm SFT}=\partial^2 \tilde{L},\quad \tilde L=L+{d-2\over 2d}\partial^2 Y\ .
\eea
The stress-energy tensor $T_{\mu\nu}^{SFT}$ is invariant under $S$, while the ``improvement scalar'' $\tilde{L}$ can not be invariant, $[S,\tilde L]=X \neq 0$. Otherwise $S$ would commute with $T_{\mu\nu}^{CFT}$ and consequently with all conformal generators, which is only possible if the SFT  is in fact conformal. Acting by $S$ on both sides of \eqref{trace} implies that $X$ satisfies the free field equation 
\bea
\label{freescalar}
\partial^2 X=0\ .
\eea
In other words the SFT in question is a free field theory. 
This can be seen in the following way. First we normalize $X$ such that the two-point function assumes the canonical form $\langle X(x)X(0)\rangle =x^{2-d}$ and consider the four-point function $\langle XXXX\rangle$. We can write it as follows
\bea
\label{fourp}
\langle X X X X \rangle &=& {1 \over (x_{13}^2 x_{24}^2)^{d/2 - 1}} \left( 1 + u^{-{d-2 \over 2}} + v^{-{d-2 \over 2}} + f_{c}(u,v) \right) , \\
u &=& {x_{12}^2 x_{34}^2 \over x_{13}^2 x_{24}^2},~~~ v= {x_{14}^2 x_{23}^2 \over x_{13}^2 x_{24}^2} .
\eea
where we separated the contribution of the unit operator in each channel and the connected piece $f_{c}(u,v)$. 
Equation \eqref{freescalar} applied at the point $x_1$ takes the following form
\bea
\label{freescalarCR}
\pa_u \pa_v f_c(u,v) - \left({d \over 2} + u \pa_u + v \pa_v \right) (\pa_u + \pa_v) f_c (u,v)=0\ .
\eea

To understand the implications \eqref{freescalarCR} it is instructive to look at the light-cone limit $u \to 0$, $v-{\rm fixed}$. On general grounds the correlation function in this limit takes the form of an expansion in $u^{{\tau - (d-2) \over 2}}$ where $\tau = \Delta - s$ is the twist of the operator that appears in the OPE. Since we isolated the contribution of the unit operator the expansion of $f_c (u,v)$ starts from $\tau > 0$. Plugging the $u$-expansion  into \eqref{freescalarCR},  one can show 
that 
\bea
\label{sol}
f_c(u,v) = \sum_{n=0}^{\infty} u^n f_n (v)\ ,
\eea
where $n$ is an integer and \eqref{freescalarCR} relates $f_n (v)$ to $f_{n-1}(v)$.

This implies that only operators with twist 
\bea
\label{twist}
\tau = d-2 + 2 n 
\eea
can appear in the OPE of $X X$. Let us focus on the contribution of operators with $n=0$ since these correspond to conserved currents. To better understand these we can consider the correlator in the limit $v \to 1$ and $u$ fixed. It is dominated by the unit operator in the dual channel and as explained in \cite{Fitzpatrick:2012yx,Komargodski:2012ek} this implies existence of an infinite set of operators with twist that approaches $2 \Delta = d-2$ for infinite spin. Together with the constraint \eqref{twist} it implies existence of an infinite number of operators with twist exactly being $d-2$. Thus, the original CFT contains an infinite number of higher spin conserved currents.

Using the arguments of \cite{Maldacena:2011jn} we can use higher spin symmetries to fix all correlators of $X$.\footnote{We have an infinite set of higher spin charges $Q_{s} = \int d x^{d-1} j_{-...-}$ built out of the conserved current $j_s$ of spin $s$. These charges act like $[Q_s, X] = \pa^{s-1}_{-} X$ which is a consequence of unitarity. We can then use an infinite set of Ward identities $\langle [Q_s, XX...X] \rangle = 0$ to fix the correlator.} 
All these correlators are the ones of the free scalar theory. Consequently correlation functions of all operators appearing in the OPE of $X$'s are also free. In particular the $n$-point correlation function of the stress-energy tensors are those of the free scalar theory.\footnote{Strictly speaking we do not show that the theory does not contain some extra set of operators such that $\langle {\cal O} X ... X \rangle = 0$. We believe this should follow from  the uniqueness of the stress-energy tensor. }

\subsection{Example: free massless scalar}\label{example}
To illustrate the general mechanism of the last section we consider the default example   of unitary relativistic scale but not conformal-invariant field theory: free massless scalar with a shift symmetry. Working in arbitrary number of spacetime  dimensions the  theory is defined by  the Lagrangian
\bea
\label{fft}
{\mathcal L}=(\partial_\mu \phi)^2\ .
\eea
This theory admits a  shift symmetry $\phi\rightarrow \phi+{\rm const}$. Let us consider a sub-algebra of operators invariant under this symmetry. Colloquially these are the operators made out of derivatives of $\phi$ but not of $\phi$  itself: $\partial_\mu \phi,\, \partial_\mu \partial_\nu \phi,\, \partial_\mu \phi\, \partial_\nu \phi,\dots$ Clearly the set of shift-invariant  operators is closed under the OPE expansion. Furthermore, 
the canonical stress-energy tensor
\bea
\label{tsft}
T^{\rm can.}_{\mu\nu}=\partial_\mu \phi\, \partial_\nu \phi-{g_{\mu\nu}\over 2}(\partial \phi)^2\ ,
\eea
is invariant under the shift symmetry as well. Hence the free theory of  shift-invariant scalar is a SFT with $T^{\rm SFT}_{\mu\nu}=T^{\rm can.}_{\mu\nu}$. In all $d>2$ this theory is not conformal as follows from a non-vanishing trace of \eqref{tsft}.

The stress-energy tensor \eqref{tsft} can be improved to a traceless form  
\bea
T^{\rm CFT}_{\mu\nu}= T^{\rm SFT}_{\mu\nu}-{(d-2)\over 4 (d-1)}(\partial_\mu \partial_\nu-g_{\mu\nu}\partial^2)\phi^2\ ,
\eea thus proving the full theory is a CFT. 

The shift symmetry  generator $S$, $[S,\phi]=1$, is an integral of a local current
\bea
S=\int d^{d-1}x\, \partial_0 \phi\ .
\eea
Crucially, $S$ does not commute with the conformal algebra: while primary $\phi$ is not invariant under $S$, the descendant $\partial_{\mu}\phi$ is. 
Accordingly  $\tilde L=-{d-2\over 4}\phi^2$ is not  invariant under the shift symmetry as well, while $X=[S,\tilde L]=-{d-2\over 2}\phi$ is indeed a free scalar satisfying \eqref{freescalar}. 

The main result of this paper  is that any SFT embedded into a unitary CFT is a theory \eqref{fft} in disguise. We could be more generic and assume $S$ to be a tensor $S=S_{\mu_1\dots \mu_{\ell}}$, rather than a scalar. Consequently, $X_{\mu_1\dots \mu_{\ell}}=[S_{\mu_1\dots \mu_{\ell}},L]$ would be a tensor as well, while $\partial^2 X_{\mu_1\dots \mu_{\ell}}=0$ would still hold. Thus, a SFT embedded into a unitary CFT could be the theory of free tensor fields, but we  do not know such  examples. Free fermions and ${d \over 2}$-forms are conformal and there is no symmetry to break conformality to scale invariance. Free $(d-2)$-forms are dual to a free scalar, i.e.~this is the example considered above. Free forms of other rank are scale-invariant without full conformal invariance, but these theories can not be embedded into a unitary CFT. It should be possible to  embed these theories into a non-unitary CFT, analogous to free Maxwell theory in $d\neq 4$, which is discussed in the next section.

\section{SFTs within non-unitary CFTs}
\label{nonU}
It is tempting to generalize our discussion to include unitary scale-invariant theories embedded into non-unitary CFTs. An example of such a theory was 
given in \cite{ElShowk:2011gz}, where it was shown that the Maxwell theory in $d\neq 4$ dimensions can be embedded into some non-unitary CFT. 
The original Maxwell theory is unitary and scale-invariant, but not conformal.
This can be seen from the stress-energy tensor
\bea
\label{nonconserved}
T^{\rm YM}_{\mu\nu}={g_{\mu\nu}\over 4}F_{\alpha\beta}^2-F_{\mu\alpha} F_{\nu\alpha}\ ,
\eea
which does not allow an improvement to the traceless form. Hence we identify $T^{\rm YM}_{\mu\nu}$ with $T^{\rm SFT}_{\mu\nu}$.

Maxwell theory could be embedded into a non-unitary CFT consisting of a free vector and a scalar ghost 
\bea
\label{fulltheory}
{\mathcal L}=-{1\over 4}F_{\alpha\beta}^2-{d-4\over 2d}(\partial_\mu A_\mu)^2-{1\over 2}\epsilon_{ab}c_a\partial^2 c_b \ .
\eea
This theory is invariant under the BRST-like symmetry 
\bea
[Q_a, A_\mu]=-i\partial_\mu c_a, \quad \{Q_a,c_b\}=-i{d-4\over d}\epsilon_{ab}\partial_\mu A_\mu\ .
\eea
Consequently we identify $S$ with the BRST generators $Q_a$ and define scale-invariant sub-sector as a sector of all $Q$-closed operators, $[Q_a,\mathcal O]=0$. One can easily check that $[Q_a,T_{\mu\nu}^{\rm YM}]=0$ as expected. 

The full theory \eqref{fulltheory} is conformal and admits a traceless conserved stress-energy tensor $T_{\mu\nu}^{\rm CFT}$.
Notice that the original stress-energy tensor $T_{\mu\nu}^{\rm SFT}=T_{\mu\nu}^{\rm YM}$ given by \eqref{nonconserved}  is not conserved, 
 $\partial_\mu^{} T_{\mu\nu}^{\rm YM}={d-4\over d}F_{\nu\alpha}\partial_\alpha(\partial A)\neq 0$. Therefore the simple relation \eqref{improv}  does not longer apply. Instead the difference $T_{\mu\nu}^{\rm CFT}-T_{\mu\nu}^{\rm SFT}$ consists of two parts, an explicitly conserved improvement term as in \eqref{improv} and a non-conserved $Q$-exact piece that balances non-conservation of $T_{\mu\nu}^{\rm YM}$,
\bea
\label{newimprov}
T_{\mu\nu}^{\rm CFT}&=&T^{\rm YM}_{\mu\nu}+\epsilon^{ab}\{Q_a,[Q_b,R_{\mu \nu}]\}+ \nonumber
\\
& &{1\over 2(d-1)}\left(\partial_\mu \partial_\alpha L_{\nu \alpha}+\partial_\nu \partial_\alpha L_{\mu \alpha} - \partial^2 L_{\mu\nu}-g_{\mu\nu} \partial_\alpha \partial_\beta L_{\alpha\beta}\right)\ .
\eea
The explicit form of spin two field $R$ and $L$ can be found in appendix A. 

The relation \eqref{newimprov} shows that $\partial_\mu T_{\mu\nu}^{\rm SFT}$ is $Q$-exact and is therefore  zero within the $Q$-closed sector. Consequently $T_{\mu\nu}^{\rm SFT}$ is a well-defined conserved stress-energy tensor of the SFT sub-sector. 

The analog of \eqref{trace} is now more complicated 
\bea
\label{newtrace}
T^{\rm SFT}_{\mu\mu}=-{1\over 2(d-1)}\left((2-d)\partial_\mu\partial_\nu L_{\mu\nu}-\partial^2 L_{\alpha\alpha}\right)-\epsilon^{ab}\{Q_a,[Q_b,R_{\mu \mu}]\}\ ,
\eea
and includes an $R_{\mu\nu}$-dependent piece. Nevertheless the rest of the logic remains the same. Since the $R_{\mu\nu}$-dependent piece is $Q$-exact it vanishes after 
acting on  \eqref{newtrace} by $S=Q_a$. Finally we obtain an analog of \eqref{freescalar}, 
\bea
\label{Xequation}
(d-2)\partial_\mu\partial_\nu X_{\mu\nu}+\partial^2 X_{\alpha\alpha}=0\ ,
\eea
where the tensor $X^a_{\mu\nu}=i[Q_a,L_{\mu\nu}]$ is given by
\bea
{X_{\mu\nu}^a\over (d-4)}=-{d-1\over d-2}(A_\mu \partial_\nu c_a+ \partial_\mu c_a A_\nu)+ {g_{\mu\nu}\, d\over 2(d-2)}(A_\alpha \partial_\alpha c_a)+ {g_{\mu\nu}(d-2)\over 2d}(\partial A)c_a\ .\ 
\eea 
As we see the main difference between the case of the non-unitary CFT considered in this section and the case of unitary CFTs considered in section \ref{CFT} is not the appearance of the extra $Q$-exact term in \eqref{newimprov}. Rather it is a non-scalar form of $X_{\mu\nu}$. Indeed, in a non-unitary theory there is no reason for spin-two part of dimension $\Delta=d-2$ operator $L_{\mu\nu}$ to vanish. As the result instead of the free field equation \eqref{freescalar} we end up with the equation \eqref{Xequation}, satisfied by some interacting spin-two operator $X_{\mu\nu}$. 

To reiterate, in the case of a SFT embedded within a non-unitary CFT there is no unitary bound on operator dimensions that rules out spin-two operator  $L_{\mu\nu}^{\perp}$ of dimension $\Delta=d-2$. Consequently, the variation of ``improvement'' operator $X_{\mu\nu}=[S,L_{\mu\nu}]$ does not have to be a scalar satisfying the free field equation. In short, we can not reach any viable conclusion about properties of scale-invariant theory embedded within a non-unitary CFT. 

Another difference with the case of section  \ref{CFT} worth mentioning is that theory \eqref{fulltheory} has two conserved traceless spin-two currents. And even if we could have shown that the theory includes a free field, it would not be enough to argue that all propagating degrees of freedom are fee.

\section{Discussion}

In this paper we discussed the possibility of embedding a unitary relativistic scale but not conformal-invariant field theory (SFT) into a unitary CFT. This question is motivated by the quest of understanding the landscape of renormalization group fixed points. With few known exceptions unitary relativistic fixed points exhibit full conformal symmetry -- a remarkable phenomenon which is not well understood. This topic goes back to 80's when
  emergence of conformal symmetry at fixed points in two dimensions was rigorously proved in \cite{Polchinski:1987dy}. Similarly, recent progress in four dimensions \cite{Luty:2012ww,Dymarsky:2013pqa, Dymarsky:2014zja, Yonekura:2014tha} implies that an SFT ought to be conformal or to be a sub-sector of a conformal theory if it can be embedded into an RG flow starting and ending at the conventional conformal fixed points. Despite  recent revival of interest  triggered by \cite{Dorigoni:2009ra}, not much is known about possibility of scale invariance without conformal invariance in $d\neq 2,4$, see  \cite{Nakayama:2013is} for a comprehensive review. Few known examples of SFTs in $d\ge 3$ are free field theories which hints there may be no interacting non-conformal unitary SFTs in any $d$. This conclusion is supported by our finding that in any number of spacetime dimensions a scale-invariant theory embedded inside a unitary CFT must be a free field theory.  Clearly, our result does not prove that non-trivial SFTs do not exist. It suggests however that non-trivial SFTs, if exist, may be rather exotic and hard to find, and would likely belong to the disconnected part of the QFT landscape.

It should be noted that our conclusion that looking for SFTs under the CFT lamppost is bound to fail rests on a number of technical assumptions. Firstly, our argument applies only to unitary CFTs. As explained in section \ref{nonU} we can not reach a viable conclusion about SFTs embedded within non-unitary CFTs, although the only known examples to date are those of free vector theories.\footnote{We expect that any theory of free forms of rank $k\neq (d-2),d/2$ in $d$ spacetime dimensions can be embedded into a non-unitary CFT by supplementing  an appropriate ghost sector, in a complete analogy with \eqref{fulltheory}.} Secondly, we  have assumed that the scale-invariant sub-sector is  a sector invariant under some symmetry $S$. This is certainly the most natural way to define a sector closed under the OPE algebra, but this is not the only possibility. 

To illustrate this idea, we consider $N$ copies of the free massless scalar, which is a non-interacting CFT. This theory admits $\R^N$ group of shift symmetries with the invariant sector being an SFT. This SFT consists of $N$ copies of scale-invariant theory of free massless scalar with shift symmetry considered in section \ref{example}.  Now, let us consider the $O(N)$-invariant sector of both the SFT in question and the enveloping CFT. The resulting theories are well-defined, and still the $O(N)$-invariant SFT is a sub-sector of the $O(N)$-invariant CFT. But there is no appropriate symmetry $S$ defined within the $O(N)$-invariant CFT, which could be used to define $O(N)$-invariant SFT through $[S,\mathcal O]=0$. Indeed, the shift symmetry $\R^N$ does not commute with $O(N)$ symmetry and hence can not be defines as a symmetry acting within $O(N)$-invariant CFT. This example is  interesting in the context of holography since the singlet sector of the $O(N)$ model is described by the Vasiliev theory of higher spin fields in AdS  \cite{Vasiliev:1999ba, Klebanov:2002ja}. Describing the gravity dual of the $O(N)$-invariant SFT from above is an interesting problem which deserves further investigation.

\vspace{1.cm}
\noindent  We would like to thank  Daniel Jafferis for stimulating discussions. 
AD gratefully acknowledges support from the grant RFBR 15-02-02092.

\section{Appendix A}
In this appendix we collect some useful formulae pertaining to the embedding of Maxwell theory in $d\neq 4$ dimensions into non-unitary CFT \eqref{fulltheory}. The stress-energy tensor of the full theory is given by \eqref{newimprov} where the stress-energy tensor of the gauge-fields $T^{\rm YM}_{\mu\nu}$ is given by \eqref{nonconserved}, while the stress-energy tensor of ghosts $T_{\mu\nu}^{\rm ghost}$ is coming from the $Q$-exact piece
\bea
\nonumber
T_{\mu\nu}^{\rm ghost}=\epsilon^{ab}\left(
\partial_\mu c_a \partial_\nu c_b -g_{\mu\nu} \partial_\alpha c_a\partial_\alpha c_b/2\right)\ ,\qquad \\ 
\nonumber
R_{\mu\nu}=-{1\over 2}A_\mu A_\nu +{g_{\mu\nu}\over 4} A_\alpha^2-{ g_{\mu\nu}\, d\over 8(d-4)}\epsilon^{ab}c_a c_b\ ,
\eea
\bea \nonumber
\epsilon^{ab}\{Q_a,[Q_b,R_{\mu \nu}]\}=T_{\mu\nu}^{\rm ghost}+
\qquad \qquad \qquad \qquad \qquad \qquad\qquad \quad
 \\ \ \ \ {d-4\over d}\left(A_\mu \partial_\nu (\partial A)\right. + \left. A_\nu \partial_\mu (\partial A)-g_{\mu\nu}A_\alpha \partial_\alpha (\partial A)-g_{\mu\nu}(\partial A)^2/2\right)\ .\nonumber \eea
The resulting stress-energy tensor $T_{\mu\nu}^{\rm YM}+\epsilon^{ab}\{Q_a,[Q_b,R_{\mu \nu}]\}$ is conserved but not traceless and requires the conventional improvement  term 
\bea
\nonumber
{1\over 2(d-1)}\left(\partial_\mu \partial_\alpha L_{\nu \alpha}+\partial_\nu \partial_\alpha L_{\mu \alpha} - \partial^2 L_{\mu\nu}-g_{\mu\nu} \partial_\alpha \partial_\beta L_{\alpha\beta}\right)\ ,
\eea
with $L_{\mu\nu}$ given by
\bea
\nonumber
L_{\mu\nu}&=& -{(d-4)(d-1)\over (d-2)}A_\mu A_\nu +{g_{\mu\nu} (d-4) d\over 4(d-2)}A_\alpha^2-{g_{\mu\nu}(d-2)\over 4}\epsilon^{ab}c_a c_b\ .
\eea

%%%%%%%%%%%%%%%%%%%%%%%%%%%%%%%

\end{document}